\documentclass[epj]{svjour}
\usepackage{amsmath,amssymb,amsfonts,graphicx,slashed,dsfont,ulem}
\usepackage{array,multirow,booktabs,dcolumn}
\usepackage[usenames]{color}
\usepackage{enumitem}
\allowdisplaybreaks[1]
\newcommand{\reff}[1]{(\ref{#1})}
\usepackage{widetext}
\newcommand{\firstREB}[1]{#1}
\newcommand{\secondREB}[1]{#1}

\begin{document}

\title{Constraints on the chiral unitary {\boldmath$\bar KN$} amplitude from 
{\boldmath$\pi\Sigma K^+$} photoproduction data}

\author{Maxim~Mai\inst{1} 
\and Ulf-G.~Mei\ss ner\inst{1,2}}

\institute{Helmholtz--Institut f\"ur Strahlen- und Kernphysik (Theorie) and 
Bethe Center for Theoretical Physics, Universit\"at Bonn, D-53115 Bonn, Germany
\and Institut f\"ur Kernphysik, Institute for Advanced Simulation 
and J\"ulich Center for Hadron Physics, Forschungszentrum J\"ulich, D-52425 J\"ulich, Germany}

\date{Received: date / Revised version: date}

\abstract{
A chiral unitary approach for antikaon-nucleon scattering in on-shell factorization 
is studied. We find multiple sets of parameters for which the model describes all 
existing hadronic data similarly well. We confirm the two-pole structure of the $\Lambda (1405)$.
The narrow $\Lambda(1405)$ pole appears at comparable positions in the complex energy 
plane, whereas the location of the broad pole suffers  from a large  uncertainty. 
In the second step, we use a simple model for  photoproduction of $K^+\pi\Sigma$ off 
the proton and confront it with the experimental data from the CLAS collaboration. It is 
found that only a few of the hadronic solutions  allow for a consistent description of the 
CLAS data within the assumed reaction mechanism.
\PACS{{12.39.Fe}{} \and {14.20.Gk}{}} 
} 

\maketitle

\section{Introduction}

The strangeness $S=-1$ resonance $\Lambda(1405)$ 
is believed to be dynamically
generated through coupled-channel effects in the antikaon-nucleon interaction. A
further intricate feature is its two-pole structure. Within chiral unitary approaches,
which are considered to be the best tool to address the chiral SU(3) dynamics in
such type of system, the investigation of the two-pole structure was initiated
in Ref.~\cite{Oller:2000fj} and thoroughly analyzed in Ref.~\cite{Jido:2003cb},
for a review see \cite{Hyodo:2011ur}. However, the $K^-p \to MB$ (with $MB = 
K^- p, \pi^+ \Sigma^-, \pi^0\Lambda, \ldots$) data alone do not allow to pin down
the poles with good precision, as it is known since long, see e.g. \cite{Borasoy:2006sr}.
Fortunately, there are other sources of information on the antikaon-nucleon dynamics
in the strangeness $S=-1$ sector. These are:
\begin{enumerate}
\item In the last years, one of the most important experimental inputs 
to be considered comes from the measurement of the characteristics of kaonic hydrogen, 
performed in the SIDDHARTA experiment at DA$\Phi$NE \cite{Bazzi:2011zj}. It 
allows for an extraction of the $K^-p$ threshold amplitude which is a combination 
of the isospin 0 and 1 components. An upgrade of the above experiment is planed 
to measure the $Kd$ threshold amplitude which can then be related to the $\bar K N$ 
amplitude directly within a non-relativistic EFT, derived in Ref.~\cite{Mai:2014xx}.
\item The invariant mass distribution $M(\Sigma^+\pi^-\pi^+)$ from the reaction 
$K^-p\to\Sigma^+\pi^-\pi^+\pi^-$ was measured in the bubble chamber experiment 
at CERN, see Ref.~\cite{Hemingway:1984pz}.
\firstREB{In a multistep production, i.e. after a production of the $\Sigma^+(1660)$ resonance, this process also involves the $\Lambda(1405)$ as an intermediate step. However, due to the low energy 
resolution and phase space suppression at higher invariant masses the obtained invariant mass distribution for the $\pi\Sigma$ final state is only useful on a qualitative level.}
\item Very precise  data from proton-proton collisions at energies of a 
few GeV have become available from COSY \cite{Zychor:2007gf} and the 
HADES  collaboration \cite{Agakishiev:2012xk}  in the last few years.
\firstREB{Here, the energy resolution is of much better quality, while the complicated reaction mechanism, $p p \to \Sigma^\pm+\pi^\mp+K^++p$, makes it very difficult to keep the model dependence of the theoretical analysis under control.}
\item Recently, very sophisticated measurements of the reaction 
$\gamma p\to K^+\Sigma \pi$ were performed by the CLAS collaboration at JLAB, 
see Ref.~\cite{Moriya:2013eb}. There, the invariant mass distribution of all 
three $\pi\Sigma$ channels was determined in a broad energy range and with 
high resolution. Finally, from these data the spin-parity analysis of the $\Lambda(1405)$ 
was performed in Ref.~\cite{Moriya:2014kpv}. There, for the first time, the quantum numbers $J^P = 1/2^-$
were deduced from an experimental measurement directly.
\end{enumerate}
While the kaonic hydrogen data have become a benchmark in analyzing $\bar KN$ scattering,
the additional information from kaonic deuterium  is still out of reach. Also,
the information from the second and third process can at present only amount to
qualitative restrictions  on the $\bar K N$ the scattering amplitude. On the other 
hand, the recent results reported by the CLAS collaboration on photoproduction of 
the $K^+\Sigma \pi$ off the proton are about to become a new benchmark of our 
understanding of the antikaon-nucleon interaction. First theoretical analyses 
have already been performed  on the basis of a chiral unitary approach in 
Refs.~\cite{Roca:2013av,Roca:2013cca,Nakamura:2013boa}. 
In Refs~\cite{Roca:2013av,Roca:2013cca}, the authors construct a simple model for the photoproduction amplitude, 
where the mechanism for the reaction $\gamma p \to K^+ \pi\Sigma$ was decomposed 
into two parts. First, the photoproduction part, $\gamma p \to K^+ MB$, for 
the meson-baryon system $MB$ of strangeness $S=-1$ was assumed to be described 
by an energy-dependent coupling constant. Then, the final-state interaction 
$MB\to\pi\Sigma$ was adopted from a chiral unitary approach including just the leading order 
effective potential on the mass shell. Albeit the simplicity of this approach, it was 
shown by the authors that, fitting the new coupling constants and modifying the strength 
of the chiral potential in a certain range, this model yields a decent description 
of the CLAS data. Another very interesting theoretical investigation was made in 
Ref.~\cite{Nakamura:2013boa}, where the reaction $\gamma p \to K^+ \pi\Sigma$ was 
also decomposed into two parts: the final-state interaction $MB\to\pi\Sigma$ and 
photoproduction part $\gamma p \to K^+ MB$. The first part is described by the 
chiral unitary approach with the kernel of  first chiral order, similar to the 
treatment of Ref.~\cite{Roca:2013av,Roca:2013cca}. Further, the photoproduction part 
was constructed in a gauge-invariant manner, i.e. coupling the photon to the Weinberg-Tomozawa, 
the Born as well as the vector meson exchange diagrams.
The line shapes of all thee states of $\pi\Sigma$ were successfully reproduced for the
four lowest energies above the $K^+ \bar K N $ threshold. Moreover, some first comparison to the 
$K^+$ angular distribution was made there.

In this paper, we take up the challenge to combine our next-to-leading order (NLO)
approach of antikaon-nucleon scattering \cite{Mai:2012dt}
\firstREB{ in an on-shell approximation as will be explained in the next section}
with the CLAS data.
First, we construct a family of solutions that lead to a good description of the
scattering and the SIDDHARTA data. This reconfirms the two-pole structure
of the $\Lambda(1405)$. As before, we find that the location of the second pole 
in the complex energy plane is not well determined from these data alone.
Then, we address the issue how this ambiguity can be constrained from the CLAS data.
Similar to Refs.~\cite{Roca:2013av,Roca:2013cca}, we use a simple semi-phenomenological
model for the photoproduction process that combines the precise NLO description
of the hadronic scattering with a simple polynomial and energy-dependent ansatz
for the process $\gamma p \to K^+ M_i B_j$, see Eq.~(\ref{eq:photo}) below. 
The corresponding energy- and channel-dependent constants are fit to the CLAS
data. Such an ansatz is perfectly fine for extracting resonance information on
such data, see the similar analysis of pion photoproduction data in 
Ref.~\cite{Ronchen:2014cna}. However, with such an ansatz it is not possible
to get a microscopic understanding of the photoproduction mechanism. For achieving that,
one would have to work along the lines outlined e.g. in 
Refs.~\cite{Borasoy:2007ku,Mai:2012wy}. This, however, goes beyond the scope of the
present paper.

Our manuscript is organized as follows. In Sec.~\ref{sec:scat}, we revisit our 
NLO approach to antikaon-nucleon scattering and construct a set of solutions that
lead to a good description of the scattering data and also the SIDDHARTA kaonic hydrogen
results. In Sec.~\ref{sec:photo} we perform a fit to the CLAS data and show how much
these help to resolve the ambiguity in the two-pole structure of the $\Lambda (1405)$.
We end with a summary and some outlook for future work. Some technicalities are
relegated to the appendices.

\section{Antikaon-nucleon scattering}
\label{sec:scat}

\subsection{Model}

In the first step of the present analysis we wish to specify the meson-baryon scattering 
amplitude in the strangeness $S=-1$ sector. As already mentioned, the goal of the present 
study is to see whether the data on $\gamma p\to \pi \Sigma K^+$, measured by the CLAS 
collaboration, allow us to put new constraints on the antikaon-nucleon interaction. 
We will assume a simplified version of the amplitude constructed and described in detail 
in Refs.~\cite{Mai:2012wy,Bruns:2010sv}, to which we refer the reader for conceptual 
details, whereas here we only present the main features of this model and concentrate 
more on new aspects.

We start from the chiral Lagrangian at  leading (LO) and next-to-leading (NLO) order, 
where the latter was first constructed in Ref.~\cite{Krause:1990xc} and reduced 
to its minimal form in Ref.~\cite{Frink:2004ic}. The corresponding chiral potential reads
 \begin{align}\label{eqn:potential}
 V(\slashed{q}_2, \slashed{q}_1; p)=&A_{WT}(\slashed{q_1}+\slashed{q_2})
 +A_{14}(q_1\cdot q_2)+A_{57}[\slashed{q_1},\slashed{q_2}]\nonumber\\
 &+A_{M} +A_{811}\Big(\slashed{q_2}(q_1\cdot p)+\slashed{q_1}(q_2\cdot p)\Big)\,,
 \end{align} 
were the incoming and outgoing meson four-momenta are denoted by $q_1$ and $q_2$, whereas 
the overall four-momentum of the meson-baryon system is denoted by $p$. The $A_{WT}$, 
$A_{14}$, $A_{57}$, $A_{M}$ and $A_{811}$ are 10-dimensional  matrices which encode the 
coupling strengths between all 10 channels of the meson-baryon system 
for strangeness $S=-1$, i.e. $\{K^-p$, $\bar K^0 n$, $\pi^0\Lambda$, $\pi^0\Sigma^0$, 
$\pi^+\Sigma^-$, $\pi^-\Sigma^+$, $\eta\Lambda$, $\eta \Sigma^0$, $K^+\Xi^-$, $K^0\Xi^0\}$. These 
matrices  depend on the meson decay constants, the baryon mass in the chiral limit, 
the quark masses as well as 14 low-energy constants (LECs) of $SU(3)$ ChPT at NLO.
These 14 LECs serve as free parameters of the present model as they are not known 
precisely at the moment\footnote{In principle, the values of these constants could be 
determined from Lattice QCD. However, the LECs entering the above potential 
are affected by several non-trivial effects, e.g. the appearance of resonances, inelastic thresholds and so on. At the moment, it is not clear, how one can include these effects systematically.}. 
The explicit form of the matrices $A_{WT}$, $A_{14}$, ... as well as the 
values of the remaining physical constants are given in 
App.~\ref{app:couplings} and App.~\ref{app:loops}, respectively.

\begin{figure*}[t]
\includegraphics[width=\linewidth]{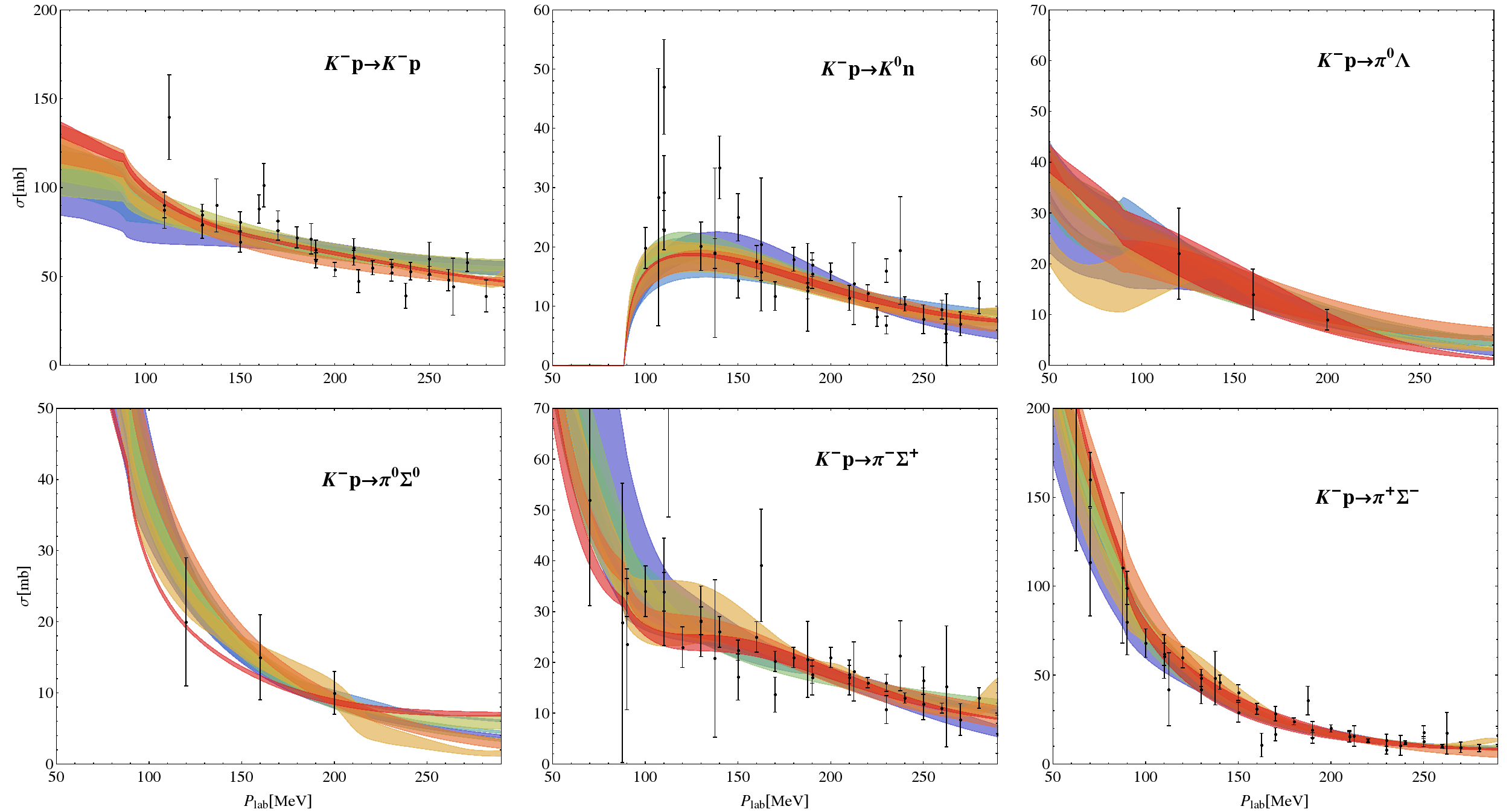}
\caption{Fit results compared to the experimental data from 
Refs.~\cite{Ciborowski:1982et,Humphrey:1962zz,Sakitt:1965kh,Watson:1963zz}. 
Different colors correspond to the eight best solutions, while the bands 
represent the $1\sigma$ uncertainty due to errors of the fit parameters. The color coding is specified in Fig.~\ref{fig:poles1}.}\label{fig:cs}
\end{figure*}

At any finite order, the strict chiral expansion of the scattering amplitude in the baryon sector is 
restricted to a certain range around the point $p^2=m_0^2$ and small momentum transfer to the baryon. 
Moreover, at any final order such a series fails in 
the vicinity of resonances such as the $\Lambda(1405)$, located just below the 
$\bar K N $ threshold. There are different ways to tackle this system. One of 
these, which became very popular over the last two decades, is the unitarization of the
chiral potential via a coupled channel Bethe-Salpeter equation (BSE), 
for NLO approaches see e.g. Ref.~\cite{Mai:2012dt,Borasoy:2005ie,Oller:2006jw,Ikeda:2011pi,Ikeda:2012au,Guo:2012vv,Borasoy:2006sr}. For the meson-baryon scattering amplitude 
$T(\slashed{q}_2, \slashed{q}_1; p)$ and the chiral potential 
$V(\slashed{q}_2, \slashed{q}_1;p)$ the integral equation to be solved reads
\begin{align}\label{eqn:BSE}
T(\slashed{q}_2, \slashed{q}_1; p)&=
  V(\slashed{q}_2, \slashed{q}_1;p) \\
  &+i\int\frac{d^d l}{(2\pi)^d}V(\slashed{q}_2, \slashed{l}; p) 
      S(\slashed{p}-\slashed{l})\Delta(l)T(\slashed{l}, \slashed{q}_1; p)\,,\nonumber
\end{align}
where $S$ and $\Delta$ represent the baryon (of mass $m$) and the meson (of mass $M$) 
propagator, respectively, and are given by $iS(\slashed{p}) = {i}/({\slashed{p}-m+i\epsilon})$
and $i\Delta(k) ={i}/({k^2-M^2+i\epsilon})$. Moreover, $T$, $V$, $S$ and $\Delta$ in the 
last expression are matrices in the channel space. The loop diagrams appearing above 
are treated using dimensional regularization and applying the usual $\overline{\rm MS}$ 
subtraction scheme in the spirit of our previous work \cite{Bruns:2010sv}, 
see App.~\ref{app:loops}. 
Note that the modified loop integrals are still scale-dependent. This scale $\mu$ reflects the 
influence of the higher-order terms not included in our potential. It is used as a 
fit parameter of our approach. To be precise, we have 6 such parameters in the isospin basis.

\begin{figure}[htb]
\includegraphics[width=\linewidth]{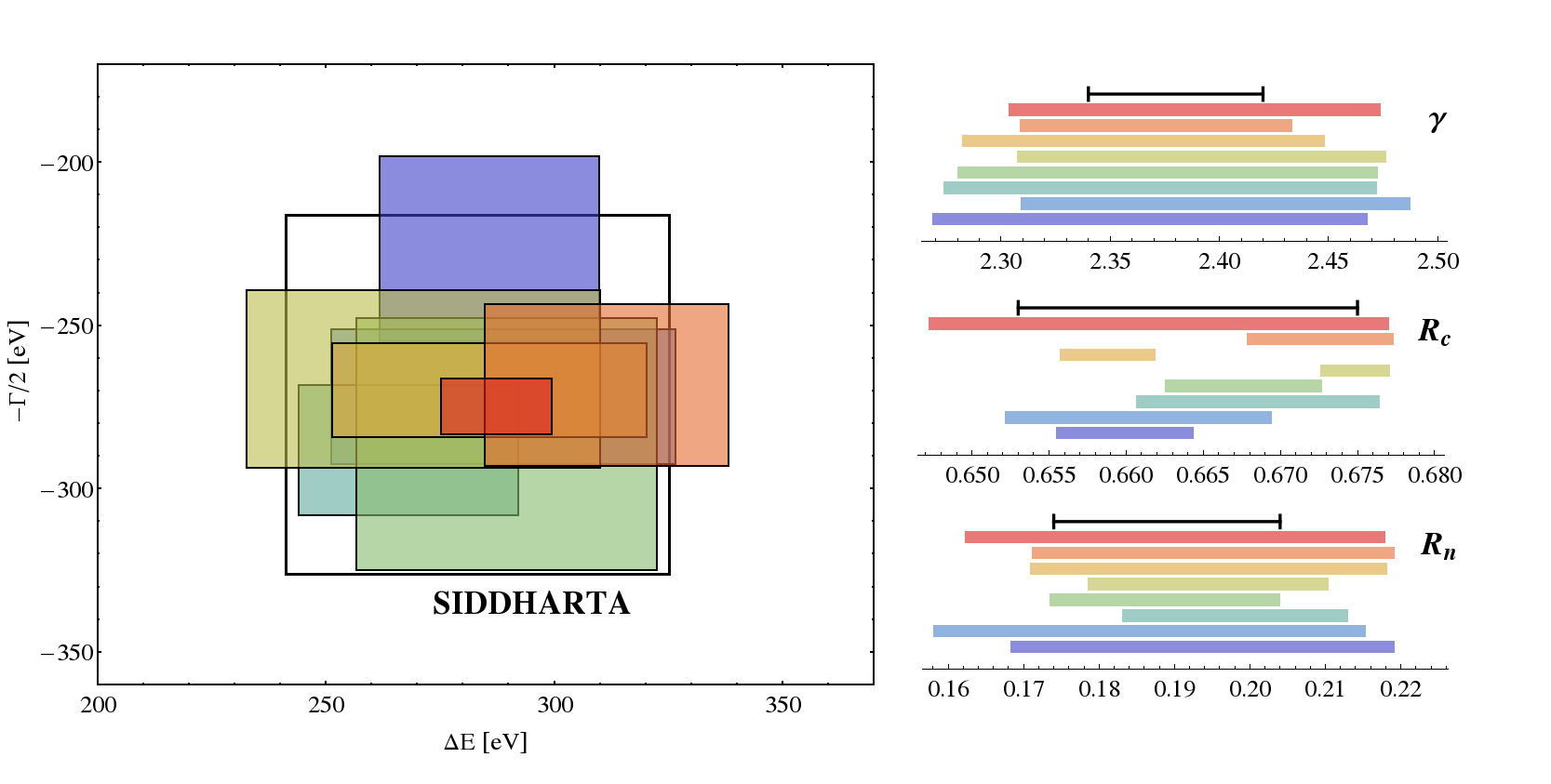}
\caption{Fit results for the threshold values as well as energy shift and 
width of kaonic hydrogen measured in \cite{Tovee:1971ga,Nowak:1978au} and 
\cite{Bazzi:2011zj}, respectively. Different colors correspond to the eight 
best solutions, while the bands represent the $1\sigma$ uncertainty  due to 
errors of the fit parameters. The color coding is specified in 
Fig.~\ref{fig:poles1} \firstREB{and the numerical values can be found in App.~\ref{app:thr}}}\label{fig:thr}
\end{figure}

\renewcommand{\baselinestretch}{1.25}
\begin{table*}[thb]
\begin{center}
\begin{tabular}{|c|cccccccc|}
\hline
Fit \# & 1&2&3&4&5&6&7&8\\
\hline
$\chi_{\rm d.o.f.}^2$ (hadronic data) &1.35 &1.14&0.99 &0.96 &1.06 &1.02
&1.15 &0.90\\
\hline
$\chi_{\rm p.p.}^2$   (CLAS data)~~ &3.18&1.94&2.56&1.77&1.90&6.11&2.93&3.14\\
\hline
\end{tabular} 
\caption{Quality of the various fits in the description of the hadronic
and the photoproduction data from CLAS. For the definition of $\chi_{\rm p.p.}^2$,
see the text. \label{tab:photo}}
\end{center}
\end{table*}

The above equation can be solved analytically if the kernel contains contact terms 
only, see Ref.~\cite{Mai:2012wy} for the corresponding solution. On the other hand, it is 
important to mention that two additional diagrams contribute already at leading 
chiral order, i.e. the $s$- and $u$-channel one-baryon exchange diagrams, also 
referred to as Born graphs. Usually, these diagrams are put on the mass shell and 
projected to a certain partial wave before including them into the BSE. Such a procedure 
seems to be quite successful from the phenomenological point of view, see 
Refs.~\cite{Ikeda:2011pi,Ikeda:2012au,Guo:2012vv} for very conclusive studies. 
The main drawback here is the loss of direct correspondence of the solution of 
Eq.~\reff{eqn:BSE} to a set of Feynman diagrams. This aspect is of great importance 
for the construction of e.g. photoproduction amplitudes in accordance with fundamental 
principles (gauge invariance) of Quantum Field Theory. Such an amplitude is constructed 
and evaluated for different channels in Refs.~\cite{Mai:2012wy,Borasoy:2007ku}. 
Therefore, we neglect the abovementioned one-baryon exchange graphs.

Up to now the only difference to our previous analysis from Ref.~\cite{Mai:2012dt} 
is the number of the meson-baryon channels included in the amplitude. In that work, 
we have also shown that once the full off-shell amplitude is constructed, one can 
easily reduce it to the on-shell solution, i.e. setting all tadpole integrals to zero. 
It appears that the double pole structure of the $\Lambda(1405)$ is preserved by 
this reduction and that the position of the two poles are changing only by about $20$~MeV 
in imaginary part\footnote{Note that this observation was made for amplitudes containing 
contact interactions only. No statement has been made there about the size of the 
contributions stemming from the inclusion of the Born graphs.}. On the other hand, 
the on-shell reduced solution of the Eq.~\reff{eqn:BSE} is much less intricate 
computationally, as it contains only two of the 20 invariant Dirac structures induced by 
the form of the kernel, see Eq.~\reff{eqn:potential}. The computational time 
therefore reduces  roughly by a factor of 30. Therefore, since we wish to explore the parameter 
space in more detail, it seems to be safe and also quite meaningful to start from 
the solution of the BSE~\reff{eqn:BSE} with the chiral potential \reff{eqn:potential} 
on the mass-shell. Once the parameter space is explored well enough we can slowly 
turn on the tadpole integrals obtaining the full off-shell solution. Such a solution 
will become a part of a more sophisticated two-meson photoproduction amplitude in 
a future publication.

\subsection{Fit procedure}

The free parameters of the present model, the low-energy constants as well as 
the regularization scales $\mu$ are adjusted  to reproduce all known experimental 
data in the meson-baryon sector. The main bulk of this data consists of the cross 
sections for the processes $K^-p\to K^-p$, $K^-p\to\bar K^0n$, $K^-p\to\pi^0\Lambda$, 
$K^-p\to\pi^+\Sigma^-$, $K^-p\to\pi^0\Sigma^0$, $K^-p\to\pi^-\Sigma^+$, see 
Refs.~\cite{Ciborowski:1982et,Humphrey:1962zz,Sakitt:1965kh,Watson:1963zz}. Here, 
only data points for the $K^-$ laboratory momentum $P_{\rm lab}<300$ MeV are considered. 
Electromagnetic effects are not included in the analysis and assumed to be negligible 
at the measured values of $P_{\rm lab}$. At the antikaon-nucleon threshold, we consider the following 
decay ratios from Refs.~\cite{Tovee:1971ga,Nowak:1978au},
\begin{align}
\gamma&=\frac{\Gamma_{K^-p\rightarrow
 \pi^+\Sigma^-}}{\Gamma_{K^-p\rightarrow \pi^-\Sigma^+}} =2.38\pm0.04\,,\nonumber\\
 R_n&=\frac{\Gamma_{K^-p\rightarrow
 \pi^0\Lambda}}{\Gamma_{K^-p\rightarrow 
 \text{neutral}}}=0.189\pm0.015\,,\nonumber\\
R_c&=\frac{\Gamma_{K^-p\rightarrow \pi^\pm\Sigma^\pm}}{\Gamma_{K^-p\rightarrow \text{inelastic}}}=0.664\pm0.011\,,
\end{align}
as well as the energy shift and width of kaonic hydrogen in the 1s state, i.e. 
$\Delta E -i\Gamma/2=(283\pm42)-i(271\pm55)$~eV from the SIDDHARTA experiment 
at DA$\Phi$NE \cite{Bazzi:2011zj}. The latter two values are related to the $K^-p$ 
scattering length via the modified Deser-type formula \cite{Meissner:2004jr}
\begin{align}
 \Delta E -i\Gamma/2=-2\alpha^3\mu^2_ca_{K^-p}
 \left[1-2a_{K^-p}\alpha\mu_c(\ln \alpha -1)\right]\,,
\end{align}
where $\alpha \simeq 1/137$ is the  fine-structure constant, $\mu_c$ is the reduced 
mass and $a_{K^-p}$ the scattering length of the $K^-p$ system.

\begin{figure*}[t]
\begin{center}
\includegraphics[width=0.8\linewidth]{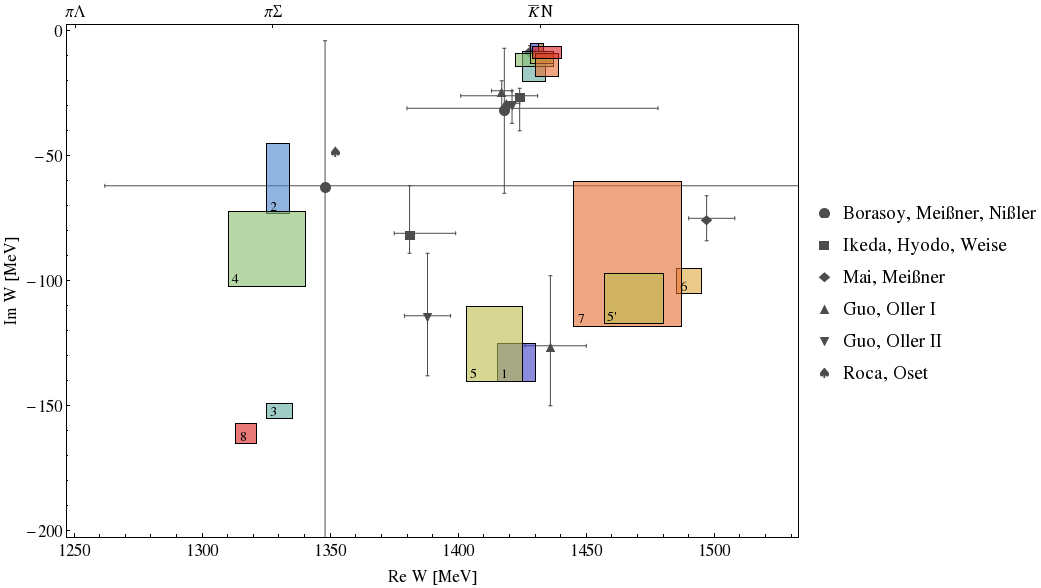}
\caption{Double pole structure of the $\Lambda(1405)$ in the complex energy
plane for the eight solutions that describe the scattering and the SIDDHARTA data.
The colors correspond to the the ones shown in Figs.~\ref{fig:cs}~and~\ref{fig:thr}. 
For easier reading,  we have labeled the second pole of these solutions by the 
corresponding fit \#, where $5$ and $5'$ denote the second pole on the second Riemann 
sheet, connected to the real axis between the $\pi\Sigma-\bar K N$ and $\bar K N-\eta\Lambda$ 
thresholds, respectively. For comparison, various results from the literature
are also shown, 
see Refs.~\cite{Borasoy:2006sr,Guo:2012vv,Ikeda:2012au,Mai:2012dt,Roca:2013av}. }
\label{fig:poles1}                                                                                \end{center}
\end{figure*}

The fit procedure was performed in two steps: First, for randomly chosen starting values of 
the free parameters (in a natural range) the fit was performed to all threshold values 
and cross section at a few momenta $P_{\rm lab}<300$ MeV. Repeating this procedure 
several thousand times,  we ended with several dozen of parameter sets that describe the
data equally well. For  each of these sets the amplitudes were analytically 
continued to the positive and negative complex plane. Thereafter, every unphysical solution, 
e.g. poles on the first Riemann sheet for ${\rm Im}(W)<200$ MeV ($W:=\sqrt{p^2}$), 
was sorted out. The remaining sets were  used in the second step as starting point of 
the fit procedure, including all  threshold and cross section data points, $\sum_i n_i=155$. 
In both steps the minimizer of the MINUIT2 \cite{MINUIT} library was applied on the
$$\chi_{\rm d.o.f.}^2:= \frac{\sum_i n_i}{(N \sum_i n_i-p)}\sum_i \frac{\chi_i^2}{n_i}~,$$ 
where $n_i$, $p$ and $N$ denote 
the number of data points for the observable $i$, the number of parameters and the overall 
number of observables, respectively. Eight best solutions were obtained by 
this two-step procedure, see  Tab.~\ref{tab:photo}, 
whereas the next best $\chi_{\rm d.o.f.}^2$ are at least one order of magnitude larger. 
Although the fit results look very promising, we would like to point out that 
there are quite a few  free parameters in the model. The latter are assumed to be 
of natural size, but not restricted otherwise. Thus, we can not exclude that there 
might be more solutions which describe the assumed experimental data equally well.

\subsection{Results}

The results of the fits are presented together with the experimental data 
in Figs.~\ref{fig:cs} and \ref{fig:thr}, where every solution is represented 
by a distinct color. The data are described equally well by all 
eight solutions, showing, however, different functional behaviour of the 
cross sections as a function of $P_{\rm lab}$. These differences are even more 
pronounced for the scattering amplitude $f_{0+}$, which is fixed model 
independently only in the ${K^-p}$ channel at the threshold by the scattering length $a_{K^-p}$.

When continued analytically to the complex $W$ plane, all eight solutions 
confirm the double pole structure of the $\Lambda(1405)$\firstREB{, see 
Fig.~\ref{fig:poles1}. There, the narrow pole lies on the Riemann sheet, connected to the real axis between the $\pi\Sigma-\bar KN$ thresholds for every solution. The second poles lie on the Riemann sheets, connected to the real axis between the following thresholds: $\pi\Sigma-\bar KN$ for solution 1, 2, 4, 5  and 8; $\pi\Lambda-\pi\Sigma$ for solution 3; $\bar K N-\eta\Lambda$ for solutions 6 and 7. Please note that the second pole of the solution 5 has a shadow pole (5' in Fig~\ref{fig:poles1}) on the Riemann sheet, connected to the real axis between $\bar K N-\eta\Lambda$ thresholds.}
The scattering amplitude is restricted around the $\bar K N$ threshold 
by the SIDDHARTA measurement quite 
strongly. Therefore, in the complex $W$ plane we observe a very stable 
behaviour of the amplitude at this energy, i.e. the position of the narrow pole 
agrees among all solutions within the $1\sigma$ parameter errors, see 
Fig.~\ref{fig:poles1}. This is in line with the findings of other 
groups~\cite{Ikeda:2012au,Borasoy:2006sr,Guo:2012vv}, i.e. one observes stability of the 
position of the narrow pole. Quantitatively, the first pole found in these models 
is located at somewhat lower energies and is slightly broader than those of our model. 
In view of the stability of the pole position, we trace this shift to the different 
treatment of the Born term contributions to the chiral potential utilized 
in Refs.~\cite{Ikeda:2012au,Borasoy:2006sr,Guo:2012vv}.

The position of the second pole is, on the other hand, less restricted. 
To be more precise, for the real part we find three clusters of these poles: 
around the $\pi\Sigma$ threshold, around the $\bar K N$ threshold as well as 
around $1470$~MeV. For several solutions there is some agreement in the positions 
of the second pole between the present analysis and the one of 
Ref.~\cite{Guo:2012vv} and of our previous work \cite{Mai:2012dt}. However, as the 
experimental data is described similarly well by all fit solutions, one  can not 
reject any of them. Thus, the distribution of poles represents the systematic 
uncertainty of the present approach.  It  appears to be quite large, but is still 
significantly smaller than the older analysis of Ref.~\cite{Borasoy:2006sr}. 
Recall that no restrictions were put on the  parameters of the model, 
except for naturalness.

\section{Photoproduction amplitude}\label{sec:photo}

\begin{figure*}[t]
\centering\includegraphics[width=0.8\textwidth]{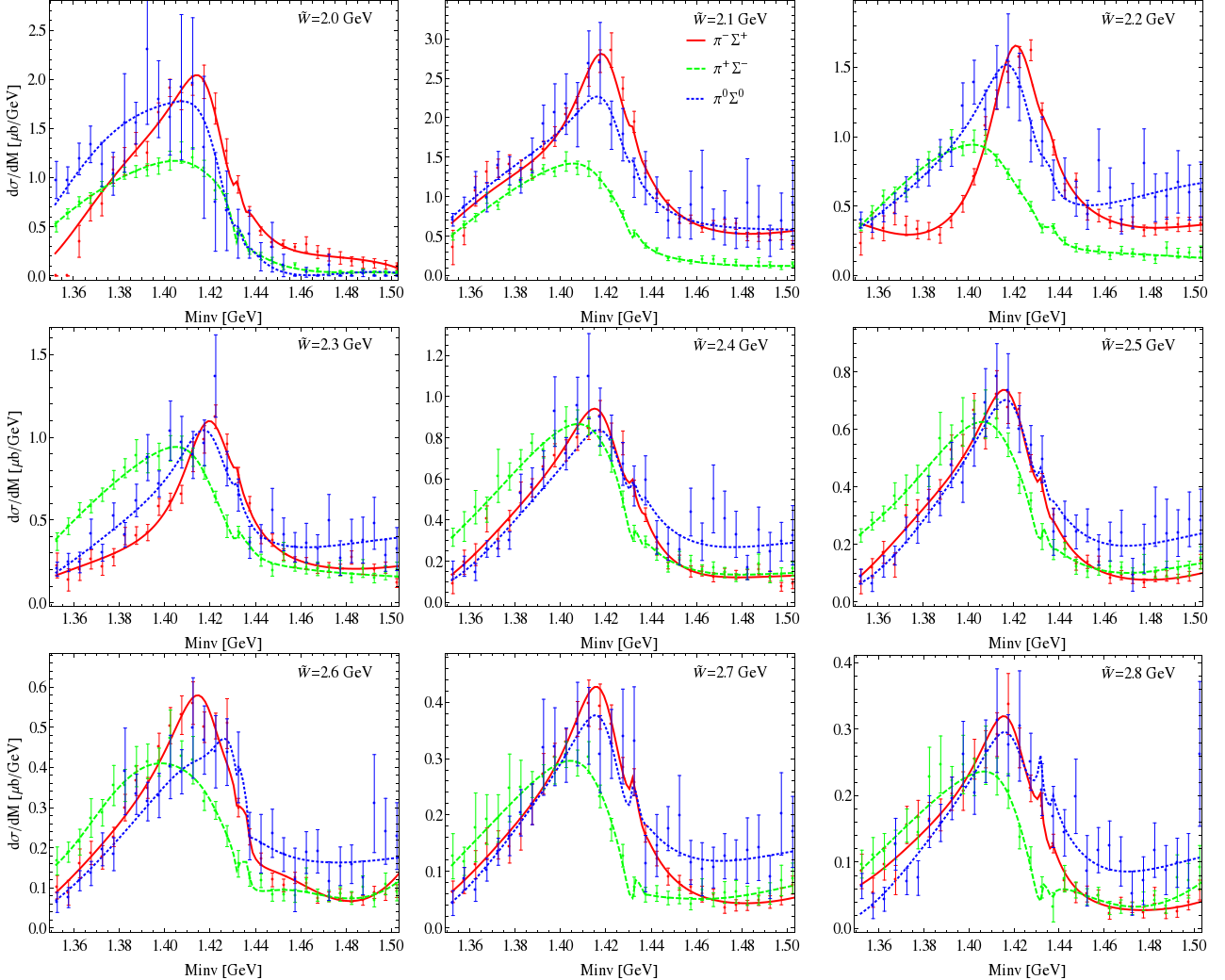}
\caption{Result of the fits to the CLAS data in all three channels $\pi^
+\Sigma^-$ (green), $\pi^-\Sigma^+$ (red) and  $\pi^0\Sigma^0$ (blue).
Correspondingly, green (dashed), red (full) and blue (dotted) lines
represent the outcome of the model for the solution \#4 in the $\pi^
+\Sigma^-$, $\pi^-\Sigma^+$ $\pi^0\Sigma^0$ channels,
respectively.}\label{fig:photobest}
\end{figure*}

\begin{figure}[t]
\centering
\includegraphics[width=0.8\linewidth]{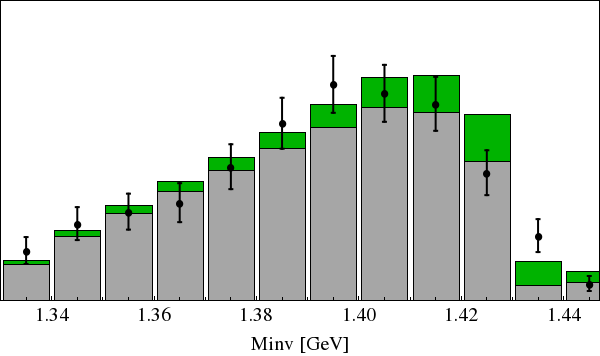}
\caption{
$\pi \Sigma$ mass distribution for the best solution~\#4 in comparison
to the Hemingway data \cite{Hemingway:1984pz}. Green bars represent 
the error bars due to propagation of $1\sigma$ error bars of the hadronic solution only.
}\label{fig:hembest}
\end{figure}

\begin{figure*}[t]
\begin{center}
\includegraphics[width=0.99\linewidth]{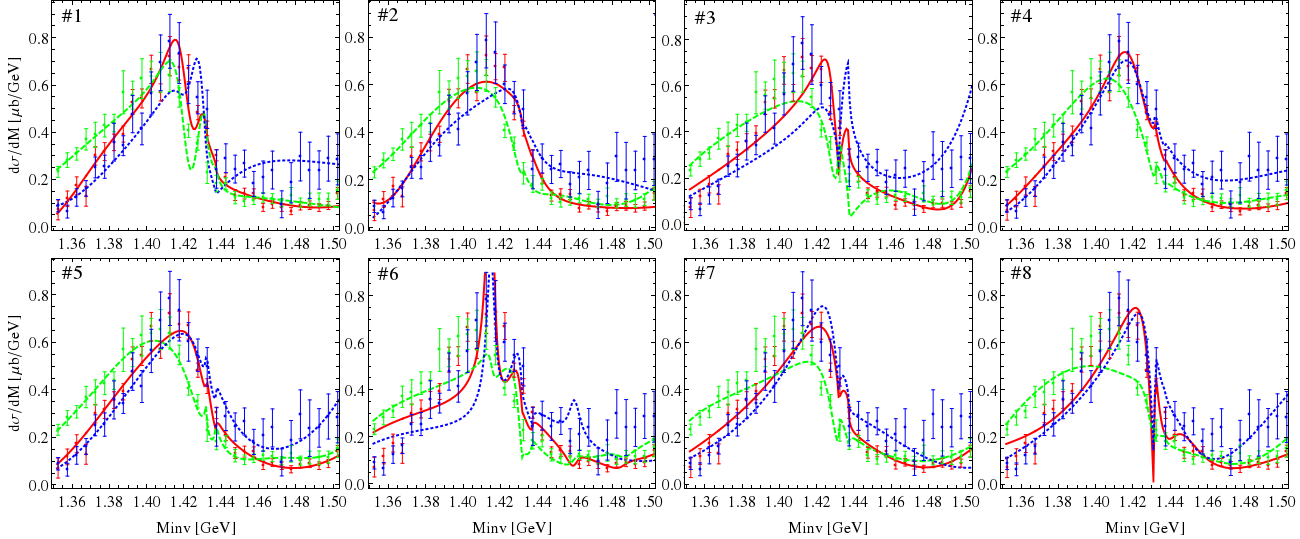}
\caption{Comparison of all solutions describing the $\pi \Sigma$ 
mass distribution at $\tilde W=2.5$ GeV in all three channels $\pi^
+\Sigma^-$ (green, dashed), $\pi^-\Sigma^+$ (full, red) and  $\pi^0\Sigma^0$ (blue, dotted).}
\label{pic:AUSSCHLUSS}
\end{center}
\end{figure*}

In the last section we have demonstrated that the present model for the 
meson-baryon interaction possess at least eight different solutions, which 
all describe the hadronic data similarly well. It is therefore of great 
importance to see how these solutions describe the photoproduction data, 
if they are considered as final-state interaction of the reaction  $\gamma p
\to K^+\Sigma \pi$. To construct such amplitudes one has to extend the framework 
for  one-meson photoproduction, described in Ref.~\cite{Borasoy:2007ku}, to the 
production amplitude including two mesons in the final state. As already mentioned 
in the beginning of the last section such amplitudes would require the full 
meson-baryon amplitude with the full off-shell dependence. For the chiral potential 
including contact interactions only, such reactions were studied in 
Refs.~\cite{Mai:2012dt,Mai:2012wy,Bruns:2010sv}, but can also be obtained from 
the present solution as argued in the last section. Then coupling the photon to every 
possible place, one obtains a gauge invariant photoproduction amplitude. While such 
a procedure is straightforward in principle, there are still many technical problems 
to overcome for the construction of such amplitudes. The goal of the present approach, 
however, is not the construction of such gauge invariant amplitudes, but to answer
the question, whether the new CLAS data allow one to rule out some of the present 
hadronic solutions. Thus, similar to  Refs.~\cite{Roca:2013av,Roca:2013cca}, we 
assume the simplest ansatz for the photoproduction amplitude 
\begin{equation}\label{eq:photo}
\mathcal M^j(\tilde W,M_{\rm inv}) 
= \sum_{i=1}^{10} C^i(\tilde W)\,G^i(M_{\rm inv})\,f_{0+}^{i,j}(M_{\rm inv})\,,
\end{equation}
where $\tilde W$ and $M_{\rm inv}$ denote the total energy of the system and the invariant mass 
of the $\pi\Sigma$ subsystem, respectively. For a specific meson-baryon channel $i$, 
the energy-dependent (and in general complex valued) constants  $C^i(\tilde W)$ describe 
the reaction mechanism of $\gamma p\to K^+M_iB_i$, where\-as the final-state interaction 
is captured by the standard H\"oh\-ler partial waves $f_{0+}$. 
For a specific meson-baryon channel $i$, the Greens function is denoted by 
$G^i(M_{\rm inv})$ and is given by the one-loop meson baryon function in dimensional 
regularization, i.e. $I_{MB}(M_{\rm inv},m_i,M_i)$ as given in App.~\ref{app:loops}.

\begin{figure*}[t]
\begin{center}
\includegraphics[width=0.8\linewidth]{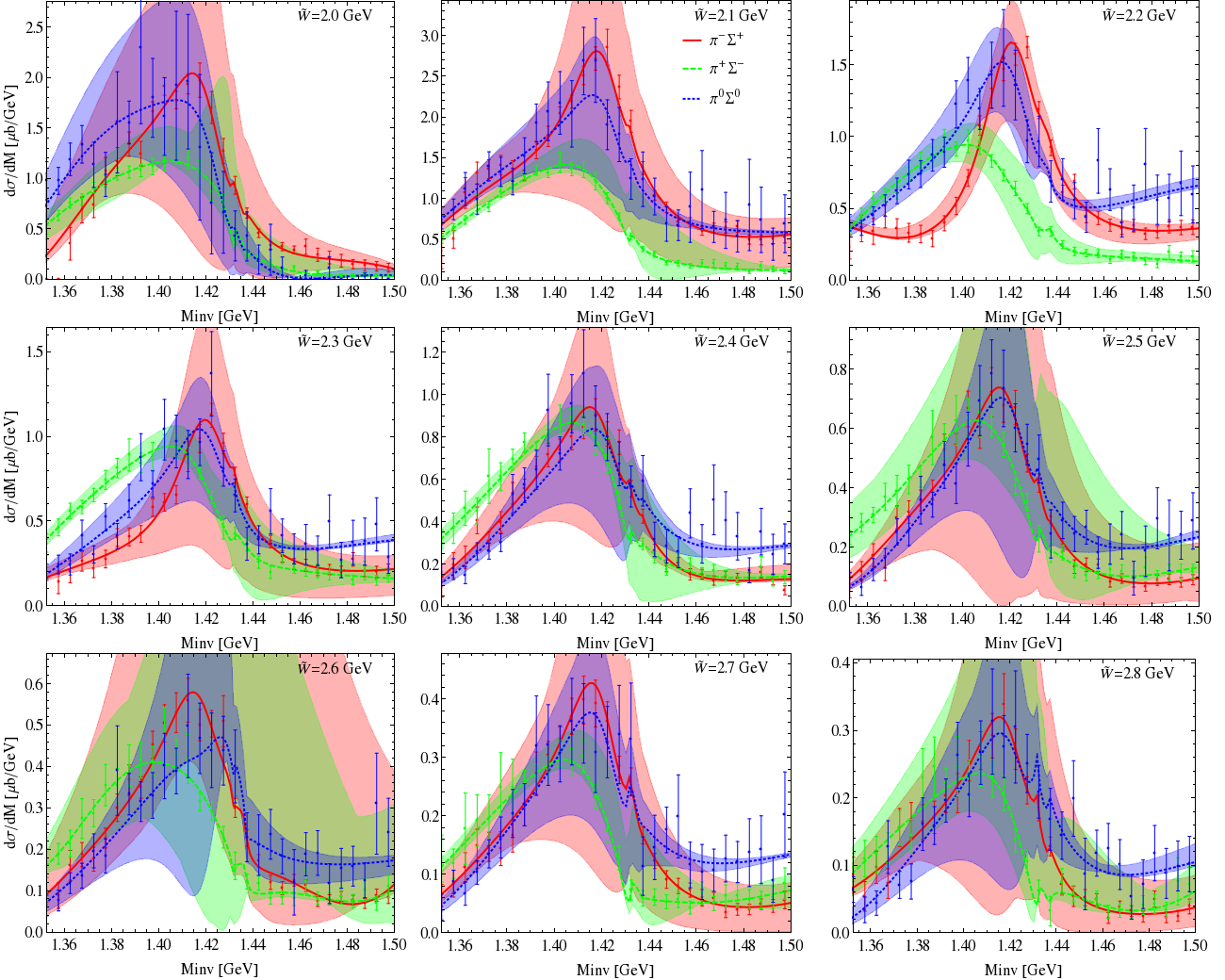}
\caption{Same as Fig.~\ref{fig:photobest}, but with error bands as described in the 
text. The color bands represent the error bars due to propagation of $1\sigma$ 
error bars of the hadronic solution which are not the $1\sigma$ error bands of the CLAS data.}
\label{fig:besterrors}
\end{center}
\end{figure*}
The regularization scales appearing in the Eq.~(\ref{eq:photo})  via the
$G^i(M_{\rm inv})$ have already been fixed in the fit to the hadronic cross sections
 and the SIDDHARTA data.  Thus, the only new parameters of the 
photoproduction amplitude are the constants $C^i(\tilde W)$ which, however, are 
quite numerous (10 for each $\tilde W$). These parameters are adjusted to 
reproduce the invariant mass distribution $d\sigma/dM_{\rm inv}(M_{\rm inv})$
for the final $\pi^+\Sigma^-$, $\pi^0 \Sigma^0$ and $\pi^-\Sigma^+$ states and
for all 9 measured total energy values $\tilde W=2.0, 2.1, .., 2.8$~GeV. The 
achieved  quality  of the photoproduction fits is listed in the third row of 
Tab.~\ref{tab:photo},
whereas the $\chi_{\rm d.o.f.}^2$ of the hadronic part are stated in the second row.
Note that for the comparison of the photoproduction fits the quantity $\chi_{\rm d.o.f.}^2$ 
is not meaningful due to the large number of generic parameters
$C_i(\tilde W)$. 
Therefore, we compare the total $\chi^2$ divided by the total
number of data points for all three $\pi\Sigma$ final states, denoted by $\chi_{\rm p.p.}^2$. 
\firstREB{For the same reason it is not meaningful to perform a global fit, minimizing the total $\chi_{\rm d.o.f.}^2$.}
It turns out that even within such a simple and flexible photoproduction
amplitude, only the solutions~\#2, \#4 and \#5 of the eight hadronic solutions
allows for a decent description of the CLAS data.
\secondREB{While the total $\chi^2$ per data point of these solutions is very close 
to each other, the next best solution has a 40\% larger total $\chi^2_{\rm p.p.}$ than 
the best one. The failure of the solutions \#1, \#3, \#6, \#7 and \#8 becomes quite 
evident in a one-to-one comparison of all eight solutions fitted to the CLAS data 
as presented in Fig.~\ref{pic:AUSSCHLUSS} for one particular cms energy chosen as a
typical example. Moreover,  the hadronic amplitudes are determined up to 1$\sigma$ 
error bands. Therefore, it is a priori not clear, whether some of the hadronic solutions 
lying within these error bands may lead to a better fit of the CLAS data. We have 
checked this explicitely, considering a large number of hadronic solutions distributed 
randomly around the central ones. For every such solution a fit to the CLAS data was 
performed independently and no significanly better fit was found to those of the 
central solution. Therefore, we consider the above exclusion principle of the 
hadronic solutions as statistically  stable.}
\renewcommand{\baselinestretch}{1.25}
\begin{table}[tbh]
\begin{center}
\begin{tabular}{|c|cc|}
\hline
solution & pole 1 & pole 2 \\
\hline
\#2  & $1434^{+2}_{-2} - i \, 10^{+2}_{-1}$ & $1330^{+4~}_{-5~} - i \, 56^{+17}_{-11}$\\
\#4  & $1429^{+8}_{-7} - i \, 12^{+2}_{-3}$ & $1325^{+15}_{-15} - i \, 90^{+12}_{-18}$\\
\hline
\end{tabular}
\end{center}
\caption{Location of the two for poles of the $\Lambda(1405)$ in the complex
energy plane (in MeV) for the two solutions that describe the scattering and the 
photoproduction data.}
\label{tab:poles}
\end{table}

The best solution is indeed \#4, which we display in Fig.~\ref{fig:photobest}.
Incidentally, it also has the lowest $\chi_{\rm d.o.f.}^2$ for the hadronic part.
This solution also gives an excellent description of the $\Sigma \pi\pi$ mass
distribution from Ref.~\cite{Hemingway:1984pz}, calculated using the method
developed in Ref.~\cite{Oller:2000fj}, c.f. Fig.~\ref{fig:hembest}.
With respect to these data,
solution \#2 is also satisfactory but \#5 is not. Therefore, the photoproduction
data combined with the scattering and the SIDDHARTA data lead to a sizable reduction 
in the ambiguity of the second pole of the $\Lambda (1405)$. In fact, the second pole 
of the surviving solutions is close to the value found in Ref.~\cite{Roca:2013av}
(which has, however, no error bars), see Fig.~\ref{fig:poles1}, and also
close to the central value of the analysis based on scattering data 
only \cite{Borasoy:2006sr}. To be precise, we give the location of the two poles
in these surviving solutions in Tab.~\ref{tab:poles}.

Given the simple phenomenological photoproduction model used here, we do not 
attempt a full-fledged error analysis of the fits to the CLAS data. However,
to get an idea about the uncertainties, we show in Fig.~\ref{fig:besterrors}
the bands generated from taking the solution within the $1 \sigma$ errors
of the hadronic amplitude for solution \#4. To be precise, these are not 
$1 \sigma$ bands for the corresponding curves but some upper limit on the
uncertainties of the fits generated from variations in the hadronic amplitude.
For a true error analysis of the photoproduction data, one has to work with
a truly microscopic model of the photoproduction amplitude, as developed 
in Refs.~\cite{Borasoy:2007ku,Mai:2012wy}.

\section{Discussion and summary}

In the present work we have utilized a chiral unitary approach to construct 
the  amplitude for antikaon-nucleon scattering at NLO in the expansion of the
interaction potential. Several sets of fit parameters 
were found to describe the data similarly well. The two-pole structure of 
the $\Lambda(1405)$ could be confirmed in each of these solutions. However, 
while the narrow pole appeared to be quite stable among these solutions, the 
broad one was found to be distributed in quite large region of the complex W-plane. 
In the second part of this work we have utilized a simple model, similar to 
the one of Ref.~\cite{Roca:2013av}, to demonstrate that one can rule out several 
hadronic solutions by demanding a good description of the CLAS photoproduction data.

We conclude that the inclusion of the CLAS data as experimental input can serve 
as a new important constraint on the antikaon-nucleon scattering amplitude. However, for  
future studies a theoretically more robust model for the two-meson photoproduction 
amplitude is required. We propose that a generalization of the one-meson photoproduction 
model, presented in Ref.~\cite{Borasoy:2007ku,Mai:2012wy}, may be the next logical 
step for this endevaour.


\section*{Acknowledgements} 
We thank Reinhard Schumacher, Ales Cieply and Peter Bruns for useful communications. This work was supported by the European
Community-Research Infrastructure Integrating Activity ``Study of Strongly Interacting 
Matter'' (acronym HadronPhysics3, Grant Agreement n. 283286) under the Seventh Framework 
Programme of EU, by the DFG and NSFC through funds provided to the Collaborative Research
Center CRC~110 ``Symmteries and the Emergence of Structure in QCD'', and by the DFG (TR~16).

\appendix

\begin{widetext}
\section{Coupling matrices}\label{app:couplings}

For the channel indices $\{b,j;i,a\}$ corresponding to the process $\phi_iB_a\rightarrow\phi_jB_b$,  the relevant coupling matrices from the leading and next-to leading order chiral Lagrangian read
\begin{align*}
A_{WT}^{b,j;i,a}=&-\frac{1}{4F_j F_i}\langle\lambda^{b\dagger}[[\lambda^{j\dagger},\lambda^{i}],\lambda^{a}]\rangle~, \\
A_{14}^{b,j;i,a}=&-\frac{2}{F_j F_i}\Big(
b_1\Big(\langle\lambda^{b\dagger}[\lambda^{j\dagger},[\lambda^{i},\lambda^{a}]]\rangle +\langle\lambda^{b\dagger}[\lambda^{i},[\lambda^{j\dagger},\lambda^{a}]]\rangle\Big)+ b_2\Big(\langle\lambda^{b\dagger}\{\lambda^{j\dagger},[\lambda^{i},\lambda^{a}]\}\rangle +\langle\lambda^{b\dagger}\{\lambda^{i},[\lambda^{j\dagger},\lambda^{a}]\}\rangle\Big)\\
& + b_3\Big(\langle\lambda^{b\dagger}\{\lambda^{j\dagger},\{\lambda^{i},\lambda^{a}\}\}\rangle +\langle\lambda^{b\dagger}\{\lambda^{i},\{\lambda^{j\dagger},\lambda^{a}\}\}\rangle\Big)
 + 2b_4 \langle\lambda^{b\dagger}\lambda^{a}\rangle \langle\lambda^{j\dagger}\lambda^{i}\rangle \Big)\,,\\
A_{57}^{b,j;i,a}=&-\frac{2}{F_j F_i}\Big(
b_5\langle\lambda^{b\dagger}[[\lambda^{j\dagger},\lambda^{i}],\lambda^{a}]\rangle
+ b_6\langle\lambda^{b\dagger}\{[\lambda^{j\dagger},\lambda^{i}],\lambda^{a}\}\rangle
+ b_7\Big(\langle\lambda^{b\dagger}\lambda^{j\dagger}\rangle \langle\lambda^{i}\lambda^{a}\rangle-\langle\lambda^{b\dagger}\lambda^{i}\rangle \langle\lambda^{a}\lambda^{j\dagger}\rangle\Big) \Big)\,,\\
A_{811}^{b,j;i,a}=&-\frac{1}{F_j F_i}\Big(
b_8\Big(\langle\lambda^{b\dagger}[\lambda^{j\dagger},[\lambda^{i},\lambda^{a}]]\rangle +\langle\lambda^{b\dagger}[\lambda^{i},[\lambda^{j\dagger},\lambda^{a}]]\rangle \Big)
+ b_9\Big(\langle\lambda^{b\dagger}[\lambda^{j\dagger},\{\lambda^{i},\lambda^{a}\}]\rangle +\langle\lambda^{b\dagger}[\lambda^{i},\{\lambda^{j\dagger},\lambda^{a}\}]\rangle\Big)\\
& + b_{10}\Big(\langle\lambda^{b\dagger}\{\lambda^{j\dagger},\{\lambda^{i},\lambda^{a}\}\}\rangle +\langle\lambda^{b\dagger}\{\lambda^{i},\{\lambda^{j\dagger},\lambda^{a}\}\}\rangle\Big)
   + 2b_{11}\langle\lambda^{b\dagger}\lambda^{a}\rangle \langle\lambda^{j\dagger}\lambda^{i}\rangle \Big)\,,\\
A_{M}^{b,j;i,a}=&-\frac{1}{2 F_j F_i}\Big(
2b_0\Big(\langle\lambda^{b\dagger}\lambda^{a}\rangle \langle[\lambda^{j\dagger}\lambda^{i}]\bar{ \mathcal{M}}\rangle\Big)
+ b_D\Big(\langle\lambda^{b\dagger}\{\{\lambda^{j\dagger},\{\bar{\mathcal{M}},\lambda^{i}\}\}\lambda^{a}\}\rangle+\langle\lambda^{b\dagger}\{\{\lambda^{i},\{\bar{\mathcal{M}},\lambda^{j\dagger}\}\},\lambda^{a}\}\rangle\Big)\\
&\hfill+ b_F\Big(\langle\lambda^{b\dagger}[\{\lambda^{j\dagger},\{\bar{\mathcal{M}},\lambda^{i}\}\},\lambda^{a}]\rangle+\langle\lambda^{b\dagger}[\{\lambda^{i},\{\bar{\mathcal{M}},\lambda^{j\dagger}\}\},\lambda^{a}]\rangle\Big)	\Big)\,,
\end{align*}
where $\{\lambda^i |i=1,...,8\}$ is a set of the $3\times 3$ channel matrices (e.g. $\phi =\phi^{i}\lambda^{i}$ for the physical meson fields) and the $F_i$ are the meson decay constants  in the respective channel. Moreover, $\bar{\mathcal{M}}$ is obtained from the quark mass matrix $\mathcal{M}$ via the Gell-Mann-Oakes-Renner relations, and given in terms of the meson masses as follows, $\bar{\mathcal{M}}=\frac{1}{2}{\rm diag}(M_{K^+}^2 - M_{K^0}^2 + M_{\pi^0}^2, M_{K^0}^2 - M_{K^+}^2 + M_{\pi^0}^2, M_{K^+}^2 + M_{K^0}^2 - M_{\pi^0}^2)\,$.

\section{Loop integrals and physical constants}\label{app:loops}

Throughout the present work we use the following numerical values for the masses and the meson decay constants (in GeV):
\begin{center}
\begin{tabular}{ccccc}
$m_{p}=0.93827$\,,     & $m_{n}=0.93956$\,,        & $m_{\Lambda}=1.11568$\,, & $m_{\Sigma^0}=1.19264$\,, & $ m_{\Sigma^-}=1.18937$\,,\\[0.1cm]
& $m_{\Sigma^+}=1.19745$\,, & $m_{\Xi^-}=1.32171$\,, & $m_{\Xi^0}=1.31486$\,, &~\\[0.1cm]
$M_{\pi^0}=0.13498$\,, & $M_{\pi^\pm}=0.13957$\,, & $M_{K^\pm}=0.49368$\,,   & $M_{\bar K^0/K^0}=0.49761$\,, & $M_{\eta}=0.54786$\,,\\[0.1cm]
& $F_\pi = 0.0924$\,, & $F_K = 0.113$\,, & $F_\eta = 1.3 F_\pi$\,. &
\end{tabular}  
\end{center}

Due to the on-shell projection of the intermediate mesons and baryons only scalar one-meson-one-baryon loop integrals appear in the context of this work. In dimensional regularization and applying applying the $\overline{\rm MS}$ 
subtraction scheme they take the following form
\begin{align*}
 I_{MB}(s,m,M)&:=\mathop{\int}_{\overline{MS}}\frac{d^dl}{(2\pi)^d}\frac{1}{l^2-M^2+i\epsilon}\frac{i}{(l-p)^2-m^2+i\epsilon}\\
              &\stackrel {d=4}{=}\frac{1}{16 \pi^2}\left(-1 +  2\log\left(\frac{m}{\mu}\right)+ \frac{M^2- m^2 + s}{s}\log\left(\frac{M}{m}\right)-2 \frac{\sqrt{\lambda(s,m^2,M^2)}}{s}\mathrm{arctanh}\left(\frac{\sqrt{\lambda(s,m^2,M^2)}}{(m + M)^2 - s}\right)\right)\,,
\end{align*}
where $\mu$ is the regularization scale and $M$ $(m)$ denotes the meson (baryon) mass. Additionally, the commonly used K\"all\'en-function $\lambda(a,b,c)=a^2+b^2+c^2-2ab-2ac-2bc$ is utilized in the above expression.

\section{Numerical results of the threshold values}\label{app:thr}
\begin{center}
\begin{tabular}{|c|cccc|}
\hline
Solution&$\Delta E -i\Gamma/2$ [eV] &$\gamma$ &$R_n$ &$R_c$ \\
\hline\hline
\#1	&$288_{-26}^{+22}-234_{-36}^{+21}  $&$2.34_{-0.07}^{+0.13 }  $&$ 0.203_{-0.035}^{+0.016} $&$ 0.658_{-0.003}^{+0.006} $\\
\#2	&$ 286_{-35}^{+40} -269_{-18}^{+24}$&$ 2.38_{-0.07}^{+0.11}  $&$ 0.190_{-0.032}^{+0.026} $&$ 0.660_{-0.007}^{+0.010} $\\
\#3	&$ 277_{-34}^{+15} -285_{-17}^{+23}$&$ 2.39_{-0.11}^{+0.09 } $&$ 0.196_{-0.013}^{+0.017} $&$ 0.671_{-0.010}^{+0.006} $\\
\#4	&$ 288_{-32}^{+34} -286_{-38}^{+39}$&$ 2.38_{-0.10}^{+0.09 } $&$ 0.191_{-0.017}^{+0.013} $&$ 0.667_{-0.005}^{+0.006} $\\
\#5	&$ 271_{-38}^{+39}-267_{-27}^{+27} $&$ 2.38_{-0.08}^{+0.09 } $&$ 0.193_{-0.015}^{+0.017} $&$ 0.675_{-0.003}^{+0.002} $\\
\#6	&$ 288_{-37}^{+32} -269_{-14}^{+15}$&$ 2.38_{-0.10}^{+0.07 } $&$ 0.195_{-0.024}^{+0.023} $&$ 0.659_{-0.003}^{+0.003} $\\
\#7	&$ 311_{-26}^{+27} -266_{-23}^{+27}$&$ 2.38_{-0.07}^{+0.05 } $&$ 0.196_{-0.025}^{+0.023} $&$ 0.671_{-0.003}^{+0.007} $\\
\#8	&$ 285_{-10}^{+14} -277_{-11}^{+7 }$&$ 2.38_{-0.08}^{+0.09 } $&$ 0.189_{-0.026}^{+0.029} $&$ 0.664_{-0.017}^{+0.013} $\\
\hline
\end{tabular}
\end{center}

\end{widetext}



\begin{thebibliography}{99}

\bibitem{Oller:2000fj}
J.~A. Oller and U.-G. Mei{\ss}ner,
Phys. Lett. B {\bf 500}  (2001) 263.
[arXiv:hep-ph/0011146].
\bibitem{Jido:2003cb}
D.~Jido, J.~A. Oller, E.~Oset, A.~Ramos, and U.-G. Mei{\ss}ner,
Nucl. Phys. A {\bf 725} (2003) 181.
[arXiv:nucl-th/0303062].
\bibitem{Hyodo:2011ur}
  T.~Hyodo and D.~Jido,
  Prog.\ Part.\ Nucl.\ Phys.\  {\bf 67} (2012) 55
  [arXiv:1104.4474 [nucl-th]].
\bibitem{Borasoy:2006sr}
  B.~Borasoy, U.-G.~Mei{\ss}ner and R.~Ni\ss ler,
  Phys.\ Rev.\ C {\bf 74} (2006) 055201,
  [hep-ph/0606108].
\bibitem{Bazzi:2011zj}
  M.~Bazzi, 
  {\it et al.},
  Phys.\ Lett.\ B {\bf 704} (2011) 113,
  [arXiv:1105.3090 [nucl-ex]].
\bibitem{Mai:2014xx}
  M.~Mai, V.~Baru, E.~Epelbaum, A.~Rusetsky
  arXiv:1411.4881 [nucl-th].
\bibitem{Hemingway:1984pz}
  R.~J.~Hemingway,
  Nucl.\ Phys.\ B {\bf 253} (1985) 742.
\bibitem{Zychor:2007gf}
  I.~Zychor
   {\it et al.},
  Phys.\ Lett.\ B {\bf 660} (2008) 167
  [arXiv:0705.1039 [nucl-ex]].
\bibitem{Agakishiev:2012xk} 
  G.~Agakishiev {\it et al.}  [HADES Collaboration],
  Phys.\ Rev.\ C {\bf 87} (2013) 025201.
\bibitem{Moriya:2013eb}
  K.~Moriya {\it et al.}  [CLAS Collaboration],
  Phys.\ Rev.\ C {\bf 87} (2013) 3,  035206
  [arXiv:1301.5000 [nucl-ex]].
\bibitem{Moriya:2014kpv} 
  K.~Moriya {\it et al.}  [CLAS Collaboration],
  Phys.\ Rev.\ Lett.\  {\bf 112}, 082004 (2014)
  [arXiv:1402.2296 [hep-ex]].
\bibitem{Roca:2013av}
  L.~Roca and E.~Oset,
  Phys.\ Rev.\ C {\bf 87} (2013) 5,  055201
  [arXiv:1301.5741 [nucl-th]].
\bibitem{Roca:2013cca}
  L.~Roca and E.~Oset,
  Phys.\ Rev.\ C {\bf 88} (2013) 5,  055206
  [arXiv:1307.5752 [nucl-th]].

\bibitem{Nakamura:2013boa}
  S.~X.~Nakamura and D.~Jido,
  PTEP {\bf 2014} (2014) 023D01
  [arXiv:1310.5768 [nucl-th]].
\bibitem{Mai:2012dt}
  M.~Mai and U.-G.~Mei{\ss}ner,
  Nucl.\ Phys.\ A {\bf 900} (2013) 51 
  [arXiv:1202.2030 [nucl-th]].
\bibitem{Ronchen:2014cna}
  D.~R\"onchen
  {\it et al.},
  Eur.\ Phys.\ J.\ A {\bf 50} (2014) 101
  [arXiv:1401.0634 [nucl-th]].
\bibitem{Borasoy:2007ku}
  B.~Borasoy, P.~C.~Bruns, U.-G.~Mei{\ss}ner and R.~Nissler,
  Eur.\ Phys.\ J.\ A {\bf 34} (2007) 161
  [arXiv:0709.3181 [nucl-th]].
\bibitem{Mai:2012wy}
  M.~Mai, P.~C.~Bruns and U.-G.~Mei{\ss}ner,
  Phys.\ Rev.\ D {\bf 86} (2012) 094033
  [arXiv:1207.4923 [nucl-th]].
\bibitem{Borasoy:2005ie}
  B.~Borasoy, R.~Ni\ss ler and W.~Weise,
  Eur.\ Phys.\ J.\ A {\bf 25} (2005) 79,
  [hep-ph/0505239].
\bibitem{Oller:2006jw}
  J.~A.~Oller,
  Eur.\ Phys.\ J.\ A {\bf 28} (2006) 63,
  [hep-ph/0603134].
\bibitem{Ikeda:2011pi}
  Y.~Ikeda, T.~Hyodo and W.~Weise,
  Phys.\ Lett.\ B {\bf 706} (2011) 63,
  [arXiv:1109.3005 [nucl-th]].
\bibitem{Ikeda:2012au}
  Y.~Ikeda, T.~Hyodo and W.~Weise,
  Nucl.\ Phys.\ A {\bf 881} (2012) 98
  [arXiv:1201.6549 [nucl-th]].
\bibitem{Guo:2012vv}
  Z.~H.~Guo and J.~A.~Oller,
  Phys.\ Rev.\ C {\bf 87} (2013) 3,  035202
  [arXiv:1210.3485 [hep-ph]].
\bibitem{Bruns:2010sv}
  P.~C.~Bruns, M.~Mai, U.-G.~Mei\ss ner,
  Phys.\ Lett.\  {\bf B697} (2011) 254,  
  [arXiv:nucl-th/1012.2233].
\bibitem{Krause:1990xc}
  A.~Krause,
  Helv.\ Phys.\ Acta {\bf 63} (1990) 3.
\bibitem{Frink:2004ic}
  M.~Frink and U.-G.~Mei{\ss}ner,
  JHEP {\bf 0407} (2004) 028,
  [arXiv:hep-lat/0404018].
\bibitem{Elvira:2014wma}
  J.~Ruiz de Elvira, C.~Ditsche, M.~Hoferichter, B.~Kubis and U.-G.~Mei\ss ner,
  EPJ Web Conf.\  {\bf 73} (2014) 05002.
\bibitem{Ciborowski:1982et}
  J.~Ciborowski, 
  {\it et al.},
  J.\ Phys.\ G {\bf 8} (1982) 13.
  
\bibitem{Humphrey:1962zz}
  W.~E.~Humphrey and R.~R.~Ross,
  Phys.\ Rev.\  {\bf 127} (1962) 1305.

\bibitem{Sakitt:1965kh}
  M.~Sakitt, T.~B.~Day, R.~G.~Glasser, N.~Seeman, J.~H.~Friedman, W.~E.~Humphrey and R.~R.~Ross,
  Phys.\ Rev.\  {\bf 139} (1965) B719.  
\bibitem{Watson:1963zz}
  M.~B.~Watson, M.~Ferro-Luzzi and R.~D.~Tripp,
  Phys.\ Rev.\  {\bf 131} (1963) 2248. 
\bibitem{Tovee:1971ga}
  D.~N.~Tovee, 
  {\it et al.},
  Nucl.\ Phys.\ B {\bf 33} (1971) 493.  
\bibitem{Nowak:1978au}
  R.~J.~Nowak, 
  {\it et al.},
  Nucl.\ Phys.\ B {\bf 139} (1978) 61. 
\bibitem{Meissner:2004jr}
  U.-G.~Mei{\ss}ner, U.~Raha and A.~Rusetsky,
  Eur.\ Phys.\ J.\ C {\bf 35} (2004) 349,
  [hep-ph/0402261]. 
\bibitem{MINUIT}
 Minuit2 released in \textbf{ROOT 5.22}/00
[http://lcgapp.cern.ch/project/cls/work-packages/mathlibs/minuit].



\end{thebibliography}
\end{document}